\documentclass[usegraphicx,usenatbib]{mn2e}
\begin{document}

\title[\textit{Chandra} Observation of Terzan 1]{A \textit{Chandra} X-ray observation of the globular cluster Terzan 1}

\author[Cackett et~al.]
 {E. M. Cackett$^1$\thanks{emc14@st-andrews.ac.uk},
 R. Wijnands$^2$,
 C. O. Heinke$^{3,4}$,
 D. Pooley$^{5,6}$,
 W. H. G. Lewin$^7$,
 \newauthor
 J. E. Grindlay$^8$,
 P. D. Edmonds$^8$,
 P. G. Jonker$^{8,9}$,
 J. M. Miller$^{10}$
\\ $^1$ School of Physics and Astronomy,
	University of St.~Andrews,
	KY16 9SS, Scotland, UK
\\ $^2$ Astronomical Institute `Anton Pannekoek',
        University of Amsterdam, Kruislaan 403, 1098 SJ,
        Amsterdam, the Netherlands
\\ $^3$	Department of Physics and Astronomy, Northwestern University,
        2145 Sheridan Road, Evanston, IL 60208, USA
\\ $^4$ Lindheimer Postdoctral Fellow
\\ $^5$ Astronomy Department, UC Berkeley, 601 Campbell Hall,
        Berkeley, CA 94720-3411, USA
\\ $^6$	Chandra Fellow
\\ $^7$ Center for Space Research, Massachusetts Institute of Technology, 77             Massachusetts Avenue, Cambridge, MA 02139, USA
\\ $^8$ Harvard-Smithsonian Center for Astrophysics, 60 Garden Street,
        Cambridge, MA 02138, USA
\\ $^9$ SRON, National Institute for Space Research, Sorbonnelaan 2,
        3584 CA Utrecht, the Netherlands
\\ $^{10}$ University of Michigan, Department of Astronomy, 500 Church Street,              Dennison 814, Ann Arbor, MI 48105 
}
\date{Accepted}
\maketitle

\begin{abstract}
We present a $\sim$19 ks \textit{Chandra} ACIS-S observation of the globular cluster Terzan 1.  Fourteen sources are detected within 1\farcm4 of the cluster center with 2 of these sources predicted to be not associated with the cluster (background AGN or foreground objects).  The neutron star X-ray transient, X1732$-$304, has previously been observed in outburst within this globular cluster with the outburst seen to last for at least 12 years. Here we find 4 sources that are consistent with the \textit{ROSAT} position for this transient, but none of the sources are fully consistent with the position of a radio source detected with the VLA that is likely associated with the transient. The most likely candidate for the quiescent counterpart of the transient has a relatively soft spectrum and an unabsorbed 0.5-10 keV luminosity of $2.6\times10^{32}$ ergs s$^{-1}$, quite typical of other quiescent neutron stars.   Assuming standard core cooling, from the quiescent flux of this source we predict long ($>400$ yr) quiescent episodes to allow the neutron star to cool. Alternatively, enhanced core cooling processes are needed to cool down the core.  However, if we do not detect the quiescent counterpart of the transient this gives an unabsorbed 0.5-10 keV luminosity upper limit of $8\times10^{31}$ ergs s$^{-1}$.  We also discuss other X-ray sources within the globular cluster.  From the estimated stellar encounter rate of this cluster we find that the number of sources we detect is significantly higher than expected by the relationship of \citet{pooley2003}.

\end{abstract}


\section{Introduction}

Neutron star X-ray transients form a subgroup of low-mass X-ray binaries.
Although usually in a quiescent state with typical X-ray luminosities of
$10^{32}$ - $10^{34}$ ergs s$^{-1}$, occasionally these systems go into
outburst where the luminosity increases to around $10^{36}$ - $10^{38}$
ergs s$^{-1}$.  These outbursts are attributed to a large increase in the mass
accretion rate onto the neutron star.
The X-ray spectra from these quiescent neutron star transients are usually
dominated by a soft component at around 1 keV, and in some cases an additional
power-law component, dominating at energies above a few keV, is present. 
The most widely accepted model used to explain the soft, thermal X-ray emission
of these transients in their quiescent states is that in which the emission is
due to the cooling of the neutron star, which has been heated during the
outbursts. In this case, the quiescent luminosity should depend on the
time-averaged accretion rate \citep{BBR98,campanaetal98a}.  However, the X-ray
emission from the cooling of the neutron star cannot directly explain the hard
power-law tail and its origin is not well understood.  Suggestions include
residual accretion down to the magnetospheric radius, or pulsar shock emission
\citep[e.g.,][]{stellaetal94,campanaetal98a,campanastella00,menoumcclintock01}.

Many quiescent neutron stars and other quiescent low-mass X-ray binaries have
been found in several globular clusters using the \textit{Chandra} and \textit{XMM-Newton} X-ray observations \citep[see][and references therein]{heinkeetal03,pooley2003}. Galactic globular clusters provide an ideal location to study these types of sources - the distance to host clusters can usually be determined more accurately than for the Galactic quiescent X-ray binaries.  The known distance and reddening allows accurate luminosities to be derived and removes the distance uncertainty from the quiescent properties.  The high incidence of compact binaries in globular clusters is likely explained by the formation of such binaries via exchange encounters in the very dense environments present \citep*[e.g.][]{verbunt87,hut91}, though the ultracompact systems may be formed via direct collisions followed by orbital decay \citep{ivanova05}.

Terzan 1 \citep{terzan66} is a globular cluster at a distance of
$5.2 \pm 0.5$ kpc and a reddening of $E(B-V) = 2.48 \pm 0.1$
\citep{ortolanietal99}.  In 1980, the \textit{Hakucho} satellite detected
X-ray bursts from a source located in Terzan 1
\citep{makishimaetal81,inoueetal81}.  Several years later, the persistent
source X1732$-$304 was detected within Terzan 1 with SL2-XRT onboard Spacelab 2
\citep{skinneretal87} and \textit{EXOSAT} \citep{warwicketal88,parmaretal89}. 
This is likely to be the same source as the bursting source detected previously
by \textit{Hakucho}.  Subsequent X-ray observations with \textit{ROSAT}
\citep{JVH95,verbuntetal95} detected the source with a similar
luminosity, between $2.0\times10^{35}$ ergs s$^{-1}$ and $1.3\times10^{36}$
ergs s$^{-1}$ \citep[see Fig.~3 of][]{GPO99}.  It is also assumed that
X1732$-$304 is the source of hard X-rays detected with the SIGMA and
ART-P telescopes \citep{borreletal96a,borreletal96b,pavlinskyetal95}.  A
possible radio counterpart of X1732$-$304 was observed with the VLA
within the \textit{ROSAT} error circles \citep{martietal98}.

A \textit{BeppoSAX} observation of X1732$-$304 in April 1999 discovered the source in a particularly low state with the X-ray intensity more than a factor of 300 lower than previous measurements \citep{GPO99}.  A more recent short ($\sim$ 3.6 ks) \textit{Chandra} HRC-I observation of Terzan 1 did not conclusively detect X1732$-$304 with a 0.5-10 keV luminosity upper limit of $(0.5-1)\times10^{33}$ ergs s$^{-1}$ depending on the assumed spectral model \citep*{WHG02}.  However, an additional X-ray source, CXOGLB J173545.6-302900, was detected by this observation.

In this paper we study the X-ray sources detected in a recent $\sim$19 ks
\textit{Chandra} ACIS-S observation of Terzan 1 and discuss possible quiescent counterparts of the neutron-star X-ray transient X1732$-$304.

\section{Observations and Analysis}

Terzan 1 was observed with the \textit{Chandra X-ray Observatory} for $\sim$19
ks on 2005 May 10. The observation was made with the Advanced CCD Imaging
Spectrometer (ACIS) with the telescope aim point on the back-illuminated S3
chip, this increases the sensitivity to low energy X-rays compared to the
front-illuminated chips. Data reduction was performed using the CIAO v3.2.2 software with the calibration database CALDB v3.1.0, provided by the \textit{Chandra} X-ray Center and following the science threads listed on the CIAO website\footnote{Available at http://cxc.harvard.edu/ciao/.}.  No background flares were found, so all available data was used in our analysis. We removed the pixel randomisation added in the standard data processing as this slightly increases the spatial resolution of the images.

\subsection{Source Detection}
\begin{figure}
\begin{center}
 \includegraphics[width = 8cm]{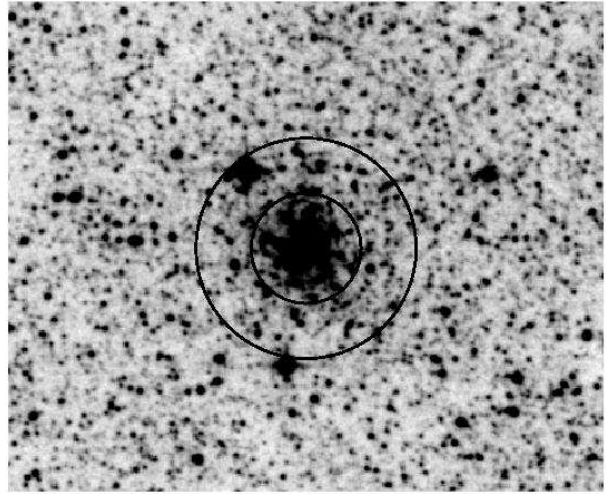}
 \caption{Digitized Sky Survey (DSS) image of Terzan 1.  This IR plate was
 originally taken with the UK Schmidt Telescope.  The inner circle indicates a
 radius of 0\farcm7 and the outer circle indicates a radius of 1\farcm4.}
 \label{fig:dss}
\end{center}
\end{figure}

\begin{figure*}
\begin{center}
 \includegraphics[width=16cm]{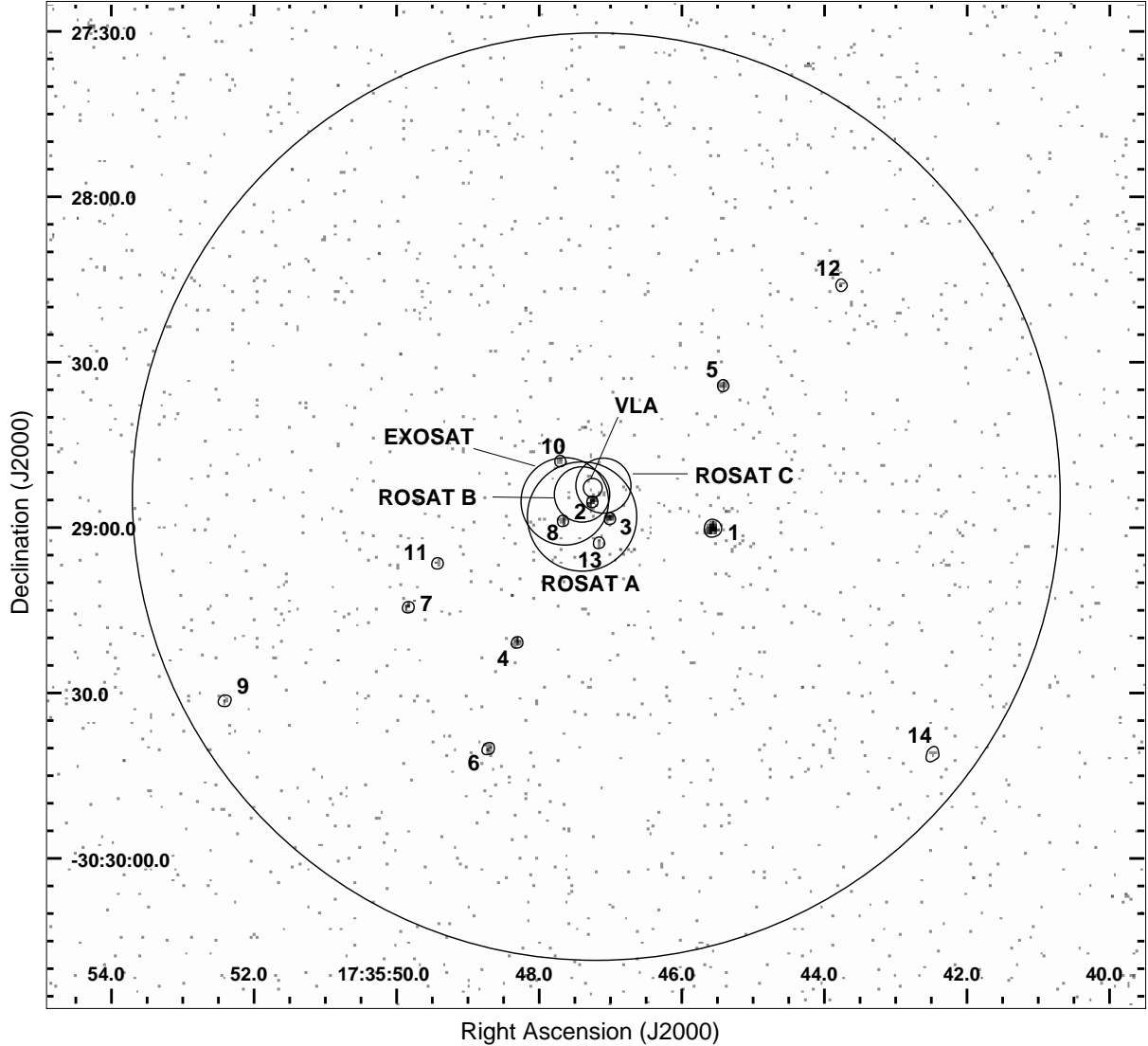}
 \caption{\textit{Chandra} ACIS-S image of the globular cluster Terzan 1 in the
 0.5-8.0 keV energy band.  The large outer circle represents the 1\farcm4 region within which we analyse the sources.  The 14 sources within this region are labelled (in order of decreasing counts in the 0.5-8.0 keV band) and the extraction regions overlaid.  Also shown are the 1$\sigma$ error circles for three \textit{ROSAT} pointings (using positions for pointings A and B from \citet{JVH95} and the updated position for pointing C from \citet{verbunt05}), the 90\% confidence level radius for the {\it EXOSAT} observation \citep{parmaretal89}, and the 90\% confidence level radius of the radio source detected with the VLA \citep{martietal98}.  We predict that 2 of the detected sources are not associated with the cluster.}
 \label{fig:sources}
 \end{center}
\end{figure*}

The CIAO tool \verb1wavdetect1 \citep{freeman02} was used to detect the sources.
We detected the sources in the 0.5-8.0 keV energy band. A detection threshold was chosen to give $\sim$1 false source over the S3 chip.  We detect 39 sources over the entire S3 chip with 4 detected counts or greater, after the addition of one extra source close to the cluster centre that was missed by \verb1wavdetect1 yet clearly a source.  Analysing the sources that are within the globular cluster half-mass radius allows a good balance between including most of the cluster sources and minimising the number of nearby objects or background AGN \citep[see Fig. 1 from][for example]{pooley02b}. \citet{trager95} define the half-mass radius for this cluster to be 3.82 arcmin, however, they note that their measurement of this value is `particularly unreliable'.  As this value for the half-mass radius is particularly large, and given that the majority of X-ray binaries in the cluster will be concentrated at the centre we chose a smaller region of size 1\farcm4 within which to analyse the X-ray sources (Fig.\ref{fig:dss} shows a DSS image of Terzan 1).  Within this 1\farcm4 region we find 14 sources.  The \textit{Chandra} image of the cluster is shown in Fig.~\ref{fig:sources} where all sources within the 1\farcm4 region are highlighted.   These sources are listed in descending number of detected counts in Table \ref{tab:sources}, also included in this table is a list of sources on the S3 chip that are not within 1\farcm4 of the cluster center, listed in decreasing 0.5-8 keV counts. The position of the brightest source we detect, CX1, is consistent with the position of the source CXOGLB J173545.6-302900 from the previous \textit{Chandra} HRC-I observation of Terzan 1 \citep{WHG02}. From the density of sources on the S3 chip outside of the 1\farcm4 region, we expect to find 2.2 sources not associated with the cluster (background or foreground objects) within the 1\farcm4 region (assuming the sources outside this region are not associated with the cluster).  This is in approximate agreement with the $\log N - \log S$ relationships of \citet{giacconi01}.  Thus we expect that 2 of the 14 sources we detect within 1\farcm4 are background or foreground sources.
\begin{table*}
\begin{center}
\caption{Names, positions (J2000), and background-subtracted counts in
four X-ray energy bands are given for the 14 sources detected within 1\farcm4 of the center of Terzan 1 and the sources detected on the rest of the S3 chip.  The statistical error (from wavdetect) in the positions of all the sources is $\le$ 0\farcs1, and thus the total positional error is dominated by the uncertainty in the absolute astrometry \citep[0\farcs6,][]{aldcroft00}.  We therefore estimate the 1$\sigma$ positional error for all sources to be $\sim$0\farcs6.  The
count extraction region for each source within the 1\farcm4 region is shown in
Fig.~\ref{fig:sources}.  The source CX1 was first detected (and named) by
\citet{WHG02}.} 
\label{tab:sources}
 \begin{tabular}{ccccccc}
 \hline
 & R.A. & Dec. &\multicolumn{4}{c}{Net Counts} \\
Source Name (Label) & (h m s) & ($^\circ$ $'$ $''$) &
 0.5 - 8.0 keV & 0.5 - 1.5 keV & 0.5 - 4.5 keV & 1.5 - 6.0 keV \\
(1) & (2) & (3) & (4) & (5) & (6) & (7) \\
\hline
CXOGLB J173545.6-302900 (CX1)  & 17 35 45.57 & -30 29 00.1 & 225$^{+16}_{-15}$ & 10$^{+4}_{-3}$    & 157$^{+14}_{-13}$ & 189$^{+15}_{-14}$  \\
CXOGLB J173547.2-302855 (CX2)  & 17 35 47.26 & -30 28 55.3 & 43$^{+8}_{-7}$    & 13$^{+5}_{-4}$    & 38$^{+7}_{-6}$    & 29$^{+7}_{-5}$  \\
CXOGLB J173547.0-302858 (CX3)  & 17 35 47.02 & -30 28 58.4 & 31$^{+7}_{-6}$    & 15$^{+5}_{-4}$    & 31$^{+7}_{-6}$    & 16$^{+5}_{-4}$  \\
CXOGLB J173548.3-302920 (CX4)  & 17 35 48.31 & -30 29 20.8 & 29$^{+6}_{-5}$    & 27$^{+6}_{-5}$    & 28$^{+6}_{-5}$    & 1$^{+2}_{-1}$   \\
CXOGLB J173545.4-302834 (CX5)  & 17 35 45.42 & -30 28 34.2 & 19$^{+5}_{-4}$    & 3$^{+3}_{-2}$     & 17$^{+5}_{-4}$    & 15$^{+5}_{-4}$  \\
CXOGLB J173548.7-302940 (CX6)  & 17 35 48.71 & -30 29 40.0 & 17$^{+5}_{-4}$    & 0$^{+2}$          & 11$^{+4}_{-3}$    & 15$^{+5}_{-4}$  \\
CXOGLB J173549.8-302914 (CX7)  & 17 35 49.84 & -30 29 14.4 & 16$^{+5}_{-4}$    & 1$^{+2}_{-1}$     & 13$^{+5}_{-4}$    & 15$^{+5}_{-4}$  \\
CXOGLB J173547.6-302858 (CX8)  & 17 35 47.67 & -30 28 58.7 & 15$^{+5}_{-4}$    & 1$^{+2}_{-1}$     & 13$^{+5}_{-4}$    & 14$^{+5}_{-4}$  \\
CXOGLB J173552.4-302931 (CX9)  & 17 35 52.41 & -30 29 31.3 & 7$^{+4}_{-3}$     & 0$^{+2}$          & 7$^{+4}_{-3}$     & 7$^{+4}_{-3}$   \\
CXOGLB J173547.7-302847 (CX10) & 17 35 47.71 & -30 28 47.9 & 6$^{+4}_{-2}$     & 1$^{+2}_{-1}$     & 5$^{+3}_{-2}$     & 4$^{+3}_{-2}$   \\
CXOGLB J173549.4-302906 (CX11) & 17 35 49.42 & -30 29 06.4 & 5$^{+3}_{-2}$     & 1$^{+2}_{-1}$     & 4$^{+3}_{-2}$     & 3$^{+3}_{-2}$   \\
CXOGLB J173543.7-302816 (CX12) & 17 35 43.77 & -30 28 16.0 & 5$^{+3}_{-2}$     & 1$^{+2}_{-1}$     & 5$^{+3}_{-2}$     & 4$^{+3}_{-2}$   \\
CXOGLB J173547.1-302902 (CX13) & 17 35 47.17 & -30 29 02.7 & 4$^{+3}_{-2}$     & 0$^{+2}$          & 4$^{+3}_{-2}$     & 4$^{+3}_{-2}$   \\
CXOGLB J173542.4-302941 (CX14) & 17 35 42.49 & -30 29 41.0 & 4$^{+3}_{-2}$     & 1$^{+2}_{-1}$     & 4$^{+3}_{-2}$     & 3$^{+3}_{-2}$   \\
Sources likely not associated with Terzan 1: & & & & & & \\
CXOU J173538.1-303032 & 17 35 38.13 & -30 30 32.1 & 128$^{+12}_{-11}$ & 2$^{+3}_{-1}$     & 100$^{+11}_{-10}$ & 121$^{+12}_{-11}$   \\
CXOU J173548.1-303420 & 17 35 48.13 & -30 34 20.2 & 74$^{+10}_{-9}$   & 66$^{+9}_{-8}$    & 73$^{+10}_{-9}$   & 8$^{+4}_{-3}$       \\
CXOU J173553.1-303433 & 17 35 53.12 & -30 34 33.2 & 56$^{+9}_{-8}$    & 46$^{+8}_{-7}$    & 57$^{+9}_{-8}$    & 10$^{+5}_{-3}$ \\
CXOU J173545.6-303204 & 17 35 45.61 & -30 32 04.9 & 45$^{+8}_{-7}$    & 37$^{+7}_{-6}$    & 44$^{+8}_{-7}$    & 8$^{+4}_{-3}$  \\
CXOU J173536.4-302917 & 17 35 36.49 & -30 29 17.9 & 31$^{+7}_{-6}$    & 5$^{+3}_{-2}$     & 30$^{+7}_{-5}$    & 26$^{+6}_{-5}$ \\
CXOU J173535.9-303143 & 17 35 35.94 & -30 31 43.4 & 30$^{+7}_{-6}$    & 2$^{+3}_{-1}$     & 21$^{+6}_{-5}$    & 25$^{+6}_{-5}$ \\
CXOU J173547.1-303426 & 17 35 47.16 & -30 34 26.1 & 27$^{+7}_{-5}$    & 7$^{+4}_{-3}$     & 24$^{+6}_{-5}$    & 21$^{+6}_{-5}$ \\
CXOU J173557.2-302839 & 17 35 57.27 & -30 28 39.2 & 19$^{+5}_{-4}$    & 3$^{+3}_{-2}$     & 15$^{+5}_{-4}$    & 15$^{+5}_{-4}$ \\
CXOU J173559.8-303016 & 17 35 59.88 & -30 30 16.8 & 17$^{+5}_{-4}$    & 5$^{+3}_{-2}$     & 17$^{+5}_{-4}$    & 12$^{+5}_{-3}$ \\
CXOU J173547.7-303332 & 17 35 45.78 & -30 33 32.5 & 17$^{+5}_{-4}$    & 1$^{+2}_{-1}$     & 12$^{+5}_{-4}$    & 14$^{+5}_{-4}$ \\
CXOU J173534.2-302715 & 17 35 34.28 & -30 27 15.3 & 13$^{+5}_{-4}$    & 1$^{+2}_{-1}$     & 10$^{+4}_{-3}$    & 11$^{+4}_{-3}$ \\
CXOU J173533.7-303002 & 17 35 33.79 & -30 30 02.5 & 13$^{+5}_{-4}$    & 1$^{+2}_{-1}$     & 7$^{+4}_{-3}$     & 8$^{+4}_{-3}$  \\
CXOU J173556.0-302717 & 17 35 56.00 & -30 27 17.3 & 12$^{+5}_{-4}$    & 1$^{+2}_{-1}$     & 7$^{+4}_{-3}$     & 10$^{+4}_{-3}$ \\
CXOU J173550.1-303155 & 17 35 50.10 & -30 31 55.0 & 9$^{+4}_{-3}$     & 3$^{+3}_{-2}$     & 10$^{+4}_{-3}$    & 7$^{+4}_{-3}$  \\
CXOU J173554.2-302728 & 17 35 54.28 & -30 27 28.8 & 8$^{+4}_{-3}$     & 0$^{+2}$          & 8$^{+4}_{-3}$     & 8$^{+4}_{-3}$  \\
CXOU J173544.2-302736 & 17 35 44.22 & -30 27 36.2 & 8$^{+4}_{-3}$     & 6$^{+4}_{-2}$     & 8$^{+4}_{-3}$     & 2$^{+3}_{-1}$  \\
CXOU J173603.7-302849 & 17 36 03.72 & -30 28 49.9 & 8$^{+4}_{-3}$     & 3$^{+3}_{-2}$     & 8$^{+4}_{-3}$     & 4$^{+3}_{-2}$  \\
CXOU J173544.0-303134 & 17 35 44.07 & -30 31 34.8 & 8$^{+4}_{-3}$     & 2$^{+3}_{-1}$     & 5$^{+3}_{-2}$     & 5$^{+3}_{-2}$  \\
CXOU J173602.1-303225 & 17 36 02.11 & -30 32 25.0 & 7$^{+4}_{-3}$     & 0$^{+2}$          & 7$^{+4}_{-3}$     & 8$^{+4}_{-3}$  \\
CXOU J173535.9-302650 & 17 35 35.91 & -30 26 50.9 & 7$^{+4}_{-3}$     & 1$^{+2}_{-1}$     & 6$^{+3}_{-2}$     & 6$^{+4}_{-2}$  \\
CXOU J173539.7-303007 & 17 35 39.77 & -30 30 07.0 & 6$^{+4}_{-2}$     & 0$^{+2}$          & 4$^{+3}_{-2}$     & 5$^{+3}_{-2}$  \\
CXOU J173555.0-302721 & 17 35 55.05 & -30 27 21.9 & 5$^{+3}_{-2}$     & 1$^{+2}_{-1}$     & 5$^{+3}_{-2}$     & 4$^{+3}_{-2}$  \\
CXOU J173540.7-302734 & 17 35 40.72 & -30 27 34.0 & 5$^{+3}_{-2}$     & 0$^{+2}$          & 5$^{+3}_{-2}$     & 5$^{+3}_{-2}$  \\
CXOU J173547.8-303039 & 17 35 47.85 & -30 30 39.0 & 4$^{+3}_{-2}$     & 4$^{+3}_{-2}$     & 4$^{+3}_{-2}$     & 0$^{+2}$       \\
CXOU J173544.7-303125 & 17 35 44.76 & -30 31 25.6 & 4$^{+3}_{-2}$     & 1$^{+2}_{-1}$     & 4$^{+3}_{-2}$     & 3$^{+3}_{-2}$  \\

\hline
\end{tabular}
\end{center}
\end{table*}

\subsection{Source Extraction}

The IDL tool ACIS Extract \citep{acis_extract} was used for source photometry
and extraction of spectra.  ACIS Extract makes use of CIAO, FTOOLS, ds9 display
capability, and the TARA IDL software.  ACIS Extract constructs polygon source
extraction regions which are approximate contours of the ACIS point-spread
function (PSF).  The user specifies the fraction of the PSF to be enclosed by
the contour.  The shape of the PSF at the location of each source is calculated
using the CIAO tool \textit{mkpsf}.  For all our sources, except one, the PSF
fraction contour level was set to 90\%, evaluated at 1.5 keV.  For the brightest
source the PSF fraction was increased to 95\%.  With the chosen contour levels
there is no overlapping of source extraction regions.

\begin{figure}
\begin{center}
 \includegraphics[width=8cm]{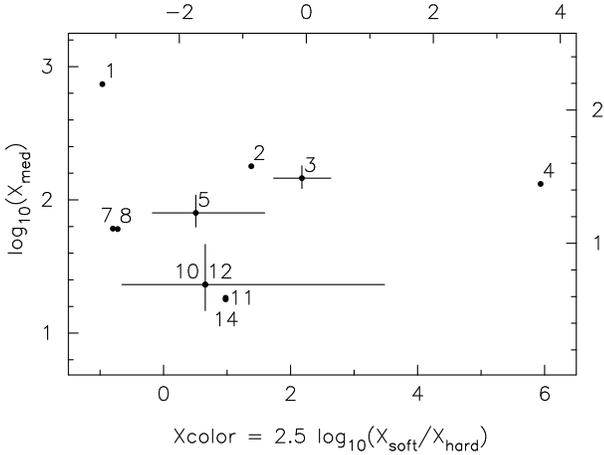}
 \caption{X-ray color magnitude diagram.  The lower and left axes show the colors and magnitudes from the absorption-corrected counts, whilst the upper and right axes show the colors and magnitudes determined from the observed counts.  We approximately correct for photoelectric absorption by shifting the data +0.69 units on the left axis and +2.25 units on the lower axis.  Absorption affects each source differently, and our correction is only an approximation based on typical spectra of globular cluster sources.  For the purposes of clarity, only a few error bars are shown.  CX6, CX9 and CX13 are not shown as they have no counts in the X$_{\mathrm{soft}}$ band.}
 \label{fig:colormag}
 \end{center}
\end{figure}

For each source a spectrum, lightcurve and event list was extracted.
The response matrix (RMF) and auxiliary response (ARF) files for each source are
constructed using the CIAO tools \textit{mkacisrmf} and \textit{mkarf}.
Background subtracted photometry was computed for each source in several
different energy bands: 0.5-1.5 keV (X$_{\mathrm{soft}}$), 0.5-4.5 keV 
(X$_{\mathrm{med}}$), and 1.5-6.0 keV (X$_{\mathrm{hard}}$).  These energy bands
were chosen to be consistent with previous globular cluster studies in the
literature \citep[e.g.][]{grindlay01a,grindlay01b,pooley02a,pooley02b,
heinkeTer5_2003,heinkeM80_2003,heinkeetal03,heinke05,bassa04}.  Table \ref{tab:sources} lists the counts in each of these bands for all the sources.
From the counts in the X$_{\mathrm{soft}}$, X$_{\mathrm{med}}$ and
X$_{\mathrm{hard}}$ bands we create an X-ray color magnitude diagram, plotting
the logarithm of the number of counts in the X$_{\mathrm{med}}$ band against the X-ray color, defined as $\mathrm{Xcolor} = 2.5\log$X$_{\mathrm{soft}}$/X$_{\mathrm{hard}}$
\citep[][though we note that Pooley et al.~do not use the factor of 2.5]{grindlay01a,grindlay01b,pooley02a,pooley02b,heinkeTer5_2003,
heinkeM80_2003,heinkeetal03,heinke05} (see Fig.~\ref{fig:colormag}). 
Photoelectric absorption reduces the number of counts detected and hence affects
the X-ray colors and magnitudes.  To allow comparison with sources in other
globular clusters this needs to be corrected for.  However, it is important to
note that absorption affects each source differently.  We determine an
approximate correction for the effects of absorption by investigating the drop
in count rate due to absorption for 3 typical spectra: a 3 keV thermal
bremsstrahlung, a 0.3 keV blackbody, and a power-law with photon index, $\Gamma = 2$. \citep[e.g.][]{pooley02a,pooley02b}.  Using PIMMS, we determine the
factor by which the count rate lowers with the inclusion of photoelectric
absorption at the column density seen for Terzan 1.  The optical reddening,
$E(B-V) = 2.48 \pm 0.1$ \citep{ortolanietal99}, is converted to a column
density $N_H = 1.36\times10^{22}$ cm$^{-2}$ \citep[$N_H/E(B-V) = 5.5\times10^{21}$ cm$^{-2}$, from][using $R = 3.1$]{predehlschmitt}.  Absorption affects the X$_{\mathrm{soft}}$ band the most, with the absorbed count rate a factor of 13.92 lower on average for the three models.  The other average correction factors are 4.85 for X$_{\mathrm{med}}$ and 1.76 for the X$_{\mathrm{hard}}$.  There is not a significant difference in the correction factors for the different models.  This leads to a shift in Xcolor of +2.25 and a shift in $\log$X$_{\mathrm{med}}$ of +0.69.  The lower and left axes of Fig.~\ref{fig:colormag} show the absorption corrected colors and magnitudes, where as the upper and right axes show the observed values. 

\subsection{Spectral Analysis}

Using \verb1XSPEC1 (v. 11.3.2) \citep{arnaud96} we fit the four brightest sources (CX1 - CX4) within the 1\farcm4 region.  We attempt to fit each of the sources with 3 different models: the neutron star atmosphere model of \citet{zavlinetal96}, a thermal bremsstrahlung model and a simple power-law.  We both fix the column density at the cluster value ($N_H = 1.36\times10^{22}$ cm$^{-2}$) and allow it to be free, using the photoelectric absorption model \verb1phabs1. Errors in the fluxes were calculated by fixing the free-parameters to their maximum and minimum value in turn, and only one at a time.  The model was then re-fitted to the data and flux values calculated.  Once this was done for every free parameter, the flux range was used to give the flux errors.

We discuss each of these sources below. Table~\ref{tab:luminosity} gives the
luminosities of the Terzan 1 sources as determined by spectral fits for CX1-3
assuming a distance to the cluster of 5.2 kpc. The luminosities for the other sources are determined by estimating the fluxes from the detected source counts using PIMMS, and assuming the cluster $N_H$ and a power-law with $\Gamma = 2.0$. A power-law with $\Gamma = 2.0$ corresponds to an absorption-corrected Xcolor$\sim0.6$, which is consistent with the majority of these sources (see Fig. \ref{fig:colormag}).
\begin{table}
\begin{center}
\caption{Luminosities for X-ray sources in Terzan 1.  For CX1-CX3 the luminosities have been calculated from the spectral fits (see text).  For the other sources, PIMMS was used to estimate the fluxes from the detected source counts, assuming the cluster $N_H$ and a power-law with $\Gamma = 2.0$.  No luminosity is given for CX4 as it is possibly a foreground object.}
\label{tab:luminosity}
\begin{tabular}{ccc}
\hline
Source & \multicolumn{2}{c}{X-ray Luminosity (ergs s$^{-1}$)} \\
 & 0.5-2.5 keV & 0.5-10 keV \\
\hline
 CX1 & $0.5^{+1.7}_{-0.3} \times 10^{33}$ & $2.0^{+1.8}_{-0.3} \times 10^{33}$  \\
CX2 & $1.9^{+1.3}_{-0.7} \times 10^{32}$ & $2.6^{+1.1}_{-0.5} \times 10^{32}$  \\
CX3 & $4.3^{+5.8}_{-2.4} \times 10^{32}$ & $4.5^{+5.7}_{-2.1} \times 10^{32}$  \\
CX5 & $4.9 \pm 1.6 \times 10^{31}$ & $8.5^{+2.4}_{-2.0} \times 10^{31}$ \\
CX6 & $8.1^{+11.7}_{-5.5} \times 10^{30}$ & $7.6^{+2.4}_{-1.9} \times 10^{31}$ \\
CX7 & $2.1^{+1.5}_{-0.9} \times 10^{31}$ & $7.2^{+2.3}_{-1.8} \times 10^{31} $ \\
CX8 & $1.3^{+1.2}_{-0.7} \times 10^{31}$ & $6.7^{+2.3}_{-1.7} \times 10^{31}$ \\
CX9  & $1.7^{+1.4}_{-0.8} \times 10^{31}$ & $3.1^{+1.7}_{-1.2} \times 10^{31}$ \\
CX10 & $3.9^{+10.0}_{-3.6} \times 10^{30}$ & $2.6^{+1.6}_{-1.1} \times 10^{31}$ \\
CX11 & $8.1^{+11.7}_{-5.5} \times 10^{30}$ & $2.2^{+1.6}_{-1.2} \times 10^{31}$ \\
CX12 & $1.3^{+1.2}_{-0.7} \times 10^{31}$ & $2.2^{+1.6}_{-1.2} \times 10^{31}$ \\
CX13 & $3.9^{+10.0}_{-3.6} \times 10^{30}$ & $1.7^{+1.5}_{-0.9} \times 10^{31}$ \\
CX14 & $1.7^{+1.4}_{-0.8} \times 10^{31}$ & $1.7^{+1.5}_{-0.9} \times 10^{31}$ \\

\hline
\end{tabular}
\end{center}
\end{table}

\subsubsection{CXOGLB J173545.6-302900 (CX1)}
\begin{figure}
\begin{center}
 \includegraphics[width=8cm]{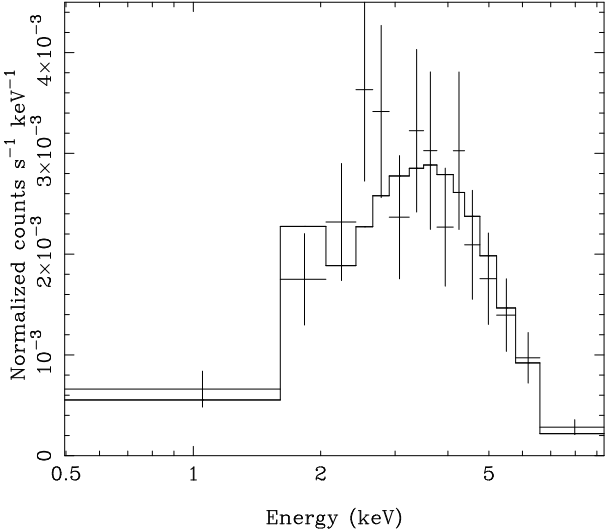}
 \caption{The spectrum of the brightest source in Terzan 1, 
 CXOGLB J173545.6-302900 (CX1).}
 \label{fig:spectrum_CX1}
\end{center}
\end{figure}

The background-subtracted spectrum was grouped to have a minimum of at least
15 counts per bin.  Both the neutron star atmosphere model and thermal bremsstrahlung model do not fit the spectrum satisfactorily with either the cluster $N_H$ or allowing it to be a free parameter. A power-law model, with absorption at the cluster value, fits with a photon index, $\Gamma = 0.2\pm0.2$ and a reduced $\chi_\nu^2$ = 0.92 for 13 degrees of freedom (dof). Allowing the photoelectric absorption to be a free parameter leads to a fit with $N_H = 2.4^{+1.8}_{-1.1} \times 10^{22}$ cm$^{-2}$, $\Gamma = 0.7^{+0.7}_{-0.6}$, and $\chi_\nu^2$ = 0.82 for 12 dof.  The photon index of the source from these power-law fits is unusually hard for globular cluster sources, and from the X-ray color-magnitude diagram it can be seen that this is the hardest source in the cluster. Only a few known globular cluster
sources have $\Gamma < 1.0$ and most tend to have $\Gamma > 1.5$.
An example of a globular cluster source that under some observing conditions
could look like CX1 is the CV X9 in 47 Tuc \citep{heinke05}. This source has very strong soft X-ray emission \citep[see Fig. 18 in][]{heinke05}, due to oxygen lines.  Above $\sim$5 keV it shows more flux than can be accounted for with simple bremsstrahlung or power-law models; possibly this is due to partial covering absorption. Observing this object only above 2 or 3 keV (for instance, if it was in Terzan 1 and heavily obscured)  would give an effective photon index of $\sim$0.3. Recently, \citet{muno2004} detected many X-ray sources in the Galactic Center that have hard spectra, with a median photon index $\Gamma = 0.7$, though many have $\Gamma \le 0$.  These authors
suggest that most of the sources are intermediate polars.  Intermediate polars
are more luminous and harder than other CVs.  The hardness of these sources
is thought to be due to local absorption (in addition to Galactic absorption)
that is partially covering the X-ray emitting region, for example by material
in the accretion flow.  We therefore model CX1 absorbed by the average cluster
absorption (using the \verb1phabs1 model, and fixing $N_H = 1.36 \times 10^{22}$
cm$^{-2}$), plus an additional
partially-covering absorption component that affects a fraction of the emitting
region (using the \verb1pcfabs1 model).  This leads to a fit with $\chi^2_{\nu}
= 0.78$ for 11 dof.  This spectral fit is shown in Fig.~\ref{fig:spectrum_CX1}.
We get a higher photon index,
with $\Gamma = 1.1^{+1.0}_{-0.9}$, with the partial absorption component having
$N_H = 3.4^{+3.8}_{-3.2} \times 10^{22}$ cm$^{-2}$ and a covering fraction of
$0.8^{+0.2}_{-0.6}$.  Thus, the partial covering absorption can explain the
extreme hardness as the photon index has increased, and the source could be a intermediate polar.  However, it is not possible to distinguish between the quality of these fits given the values of reduced $\chi^2_\nu$. 

From this model we calculate the unabsorbed 0.5-2.5 keV luminosity =
$0.5^{+1.7}_{-0.3} \times 10^{33}$ ergs s$^{-1}$ and the unabsorbed 0.5-10 keV
luminosity = $2.0^{+1.8}_{-0.3} \times 10^{33}$ ergs s$^{-1}$ (assuming a
distance of 5.2 kpc).

\subsubsection{CXOGLB J173547.2-302855 (CX2)}
As this source only has a total of 42.8 net counts in the 0.5-8 keV band, we did not group this spectrum into bins and cannot use $\chi^2$ statistics.
We therefore fit the spectrum in \verb1XSPEC1
using Cash statistics \citep{cash79} without subtracting a background spectrum
(as Cash statistics cannot be used on background subtracted spectra).  This
should introduce a minimal error as there are only 0.2 background photons in the
extraction region.  To investigate the quality of fits, the `goodness' command
is used.  This generates 10,000 Monte Carlo simulated spectra based on the
model. If the model provides a good description of the data then approximately 50\% of the simulations should have values of the Cash fit statistic that are lower than that of the best-fitting model.  We define the fit quality as the fraction of the simulated spectra with values of the Cash statistic lower than that of the data.  Both very low and very high values of the fit quality indicate that the model is not a good representation of the data.

Fitting a neutron star atmosphere model to this spectrum (with the radius = 10
km, the mass = 1.4 M$_\odot$ and the normalisation = 1/d(pc)$^2$ =
3.7$\times10^{-8}$) does not give a good fit, with a fit quality of 1.0. 
Fitting a thermal bremsstrahlung model also gives a reasonably poor fit, with a fit quality of 0.85.  An improved fit is obtained when this spectrum is fitted with a simple power-law model at the cluster $N_H$, giving a photon index,
$\Gamma = 2.5 \pm 0.6$ and a fit quality of 0.65.  Allowing the $N_H$ to be a
free parameter, gives $N_H = 1.2^{+0.8}_{-0.6} \times 10^{22}$ cm$^{-2}$,
$\Gamma = 2.3 \pm 1.0$ and a fit quality of 0.7.  Such a value of the photon index suggests that this could be a hybrid thermal plus power-law spectrum, but there are not enough counts to be conclusive.  From the X-ray color-magnitude diagram we see that this source is relatively soft and so it is possible that it could be a quiescent neutron star X-ray binary.  We determine a 0.5-2.5 keV luminosity of $1.9^{+1.3}_{-0.7} \times 10^{32}$ ergs s$^{-1}$ and a 0.5-10 keV luminosity of $2.6^{+1.1}_{-0.5} \times 10^{32}$ ergs s$^{-1}$ from the power-law fit with the cluster column density.

\subsubsection{CXOGLB J173547.0-302858 (CX3)}
This source also has too few counts to use $\chi^2$ statistics and so we again
use Cash statistics.  A simple power-law fit at the cluster column density gives
a photon index, $\Gamma = 3.9 \pm 1.2$ and a fit quality of 0.51.  Such a high
photon index strongly suggests a soft thermal spectrum, which is also indicated
by the X-ray color. A neutron star atmosphere fit (with the radius = 10
km, the mass = 1.4 M$_\odot$ and the normalisation = 1/d(pc)$^2$ =
3.7$\times10^{-8}$) at the cluster column density gives $T_{\mathrm{eff}}^{\infty} = 73.7^{+3.3}_{-3.7}$ eV with a fit quality of 0.65. Allowing the column density to be a free parameter does not greatly improve the fit, with  $N_H = 1.8^{+0.5}_{-0.4} \times 10^{22}$ cm$^{-2}$, $T_{\mathrm{eff}}^{\infty} = 78.9 \pm 6.7$ eV and a fit quality of 0.62. A thermal bremsstrahlung fit with the cluster column density gives $kT = 0.6^{+0.6}_{-0.2}$ keV with a fit quality of 0.68, where as with the column density left free we get $N_H = 1.8^{+1.3}_{-0.9}$ cm$^{-2}$ and
 $kT = 0.4^{+1.0}_{-0.2}$ keV with a fit quality of 0.77. We determine a 0.5-2.5 keV luminosity of $4.3^{+5.8}_{-2.4} \times 10^{32}$ ergs s$^{-1}$ and a 0.5-10 keV luminosity of $4.5^{+5.7}_{-2.1} \times 10^{32}$ ergs s$^{-1}$ from the power-law fit with the cluster column density.  From its spectrum, this source could be a quiescent neutron star.

\subsubsection{CXOGLB J173548.3-302920 (CX4)}
Cash statistics have also been used to analyse this source.  This source is the softest source within the cluster, with an extremely high X-ray color. There is no valid power-law fit when fixing the column density to the cluster value.  However, leaving this free we get a fit with $N_H = 0.7^{+0.5}_{-0.6} \times 10^{22}$ cm$^{-2}$, $\Gamma = 7.1^{+2.9}_{-4.3}$ and a fit quality of 0.74.  There is no good fit for either a neutron star atmosphere model or a bremsstrahlung model.  Such a soft X-ray color and spectrum, and the measured $N_H$ being much lower than the cluster value suggest that this source may not be associated with the cluster and maybe a foreground object with a lower column density.  However, there is no counterpart seen in the DSS or 2MASS images.  As this could possibly be a foreground object and the model is highly uncertain, we have not determined a luminosity for this source.

\section{Discussion}

\subsection{The neutron-star X-ray transient X1732$-$304 in quiescence}

The most likely candidate for the quiescent counterpart of the neutron-star X-ray transient X1732$-$304 is CX2.  Of the 4 sources that are within the \textit{ROSAT} error circles, it is the only one contained in all three and contained within the {\it EXOSAT} error circle (see  Fig.~\ref{fig:sources}, note that the 3 separate error circles are from the 3 separate {\it ROSAT} observations of the source).  From the X-ray color-magnitude diagram we see that this source is relatively soft and its spectrum suggests that it could be a hybrid thermal plus power-law source.  Thus, it is possible that it could be a quiescent neutron star X-ray binary.  The fact that it cannot be fit with a power-law spectrum alone is consistent with what was found by \citet{jonker04}; that the power-law fractional contribution increases for sources with luminosities decreasing from $10^{33}$ erg s$^{-1}$ to $10^{32}$ erg s$^{-1}$.
The position of CX2 is consistent with the position of 3 photons detected by \citet{WHG02} and has a flux slightly lower than their predicted upper-limit.  CX3 also has a suitably soft X-ray color and spectrum to be a possible candidate, however, its position is only fully consistent with the {\it ROSAT} A pointing and it is on the edge of the error circles for the {\it EXOSAT} and {\it ROSAT} B and C observations.  Although CX3 cannot be ruled out as the possible quiescent counterpart, CX2 is the most likely candidate.

The source CX2 is also closest to, though not fully consistent with, the position of the radio source detected with the VLA which might be the radio counterpart of the transient.  The position of this VLA source (RA: $17^h35^m47.^s 27$, Dec: $-30^{\circ}28^{\prime}52\farcs8$) is accurate to 1.7\arcsec~\citep[90\% confidence level radius,][]{martietal98}, however, the phase calibrator for these VLA observations had calibrator code C, which could mean a maximum systematic error of 0\farcs15 (J. Marti, private communication).   The offset between the radio and X-ray position is $2.5^{\prime\prime}$.  The \textit{Chandra} pointing accuracy is within the typical 0\farcs6 absolute astrometry (90\% confidence level radius) \citep{aldcroft00} as we find that the position of the X-ray source CXOGLB J173545.6-302900 (CX1) is consistent with the position of the same source in a previous \textit{Chandra} observation of the cluster, with an offset of only 0\farcs3.  Given the uncertainties in the X-ray and particularly the radio position only a small systematic offset is required for the sources to be fully consistent.

For an additional check of the accuracy of the \textit{Chandra} pointing, possible optical and infra-red counterparts were searched for.  We searched for USNO-B1.0 \citep{monet2003} point sources that are close to ACIS Extract positions for the X-ray sources on the S3 chip.  From this we get 3 optical sources that match the X-ray positions to within the 0\farcs6 absolute astrometry (90\% confidence level radius) for \textit{Chandra} \citep{aldcroft00}.  These sources are CXOU~J173544.2-302736, CXOU~J173545.6-303204 and CXOU~J173536.4-302917 and have offsets of 0\farcs07, 0\farcs39, and 0\farcs54 respectively.  The corresponding USNO-B1.0 sources are 0594-0571302, 0595-0586465, and 0594-0571303 which have positional accuracies of 0\farcs2, 0\farcs1, 0\farcs4 respectively (averaging the error in R.A. and Dec.).  Matches with the 2MASS All-Sky Catalog of Point Sources \citep{cutri2003} were also searched for.  Of the USNO-B1.0 matched sources, CXOU~J173545.6-303204 is the only one to match a 2MASS source, with an offset of 0\farcs58.  CXOU~J173550.1-303155 also has a possible 2MASS counterpart, with an offset of 0\farcs57.  The possible optical (USNO-B1.0) counterpart for this source has an offset of 0\farcs78. To check how many accidental matches we get, we apply a random offset of up to $20^{\prime\prime}$ and count the number of matches (sources found to be within 0\farcs6 of each other).  This was repeated 1000 times and the number of mean accidental matches of X-ray sources with the USNO-B1.0 catalogue was found to be $0.2 \pm 0.5$.  The 3 sources we find is significantly above this mean.

As the radio source is not seen to be fully consistent with any X-ray source there is the possibility that it is an unrelated source, for example a background AGN or quasar, in which case the radio source would likely still be present and any future radio observation might prove this.
It is also possible that we may not have detected X1732$-$304 in quiescence.  Only one photon is detected in the 0.5-8 keV band within the VLA error circle,  which could very likely be a background photon.  Assuming fewer than 5 counts were detected in VLA error circle, this gives an upper limit on the count rate of $2.7\times10^{-4}$ counts s$^{-1}$.  We use \verb1XSPEC1 to estimate an upper limit on the flux from this count rate.  Initially we fix the parameters of an absorbed neutron star atmosphere model to the appropriate cluster values expect for the effective temperature.  This parameter is then adjusted to match the required count rate of $2.7\times10^{-4}$ counts s$^{-1}$.  The unabsorbed 0.5-10 keV luminosity upper limit is then determined to be $8\times10^{31}$ ergs s$^{-1}$. Such a luminosity is slightly lower than typically seen for quiescent neutron stars \citep[e.g.][]{heinkeetal03,pooley2003} although this scenario cannot be ruled out.

Assuming that CX2 is the quiescent counterpart to X1732$-$304, it is possible to put limits on the length of time that the source will stay in quiescence.  In the deep crustal heating model of quiescent neutron star emission \citep{BBR98} the quiescent luminosity depends on the time-averaged accretion rate of the source.  Assuming standard core cooling, it is then possible to relate the quiescent flux, $F_q$, and the average flux during outburst, $\langle F_0 \rangle$, with the average time the source is in outburst, $t_0$, and quiescence, $t_q$, via $F_q \approx t_o/(t_o + t_q) \times \langle F_o \rangle/135$ \citep[e.g.,][]{wijnands2001,WHG02}.  We use the bolometric flux during outburst as estimated by \citet{WHG02} to be $1.5\times10^{-9}$ ergs cm$^{-2}$ s$^{-1}$.  To place upper limits on the thermal quiescent flux we use the neutron star atmosphere fit to the spectrum of CX2 with the column density at the cluster value.  We fix the temperature to the upper limit from the fit and determine the bolometric flux (generating a dummy response from 0.001 - 100 keV and extrapolating the model).  This gives $F_q < 3.3 \times 10^{-13}$ ergs cm$^{-2}$ s$^{-1}$.  The outburst from X1732$-$304 was seen to last at least 12~yr, so, assuming that $t_0 = 12$ yr, we get $t_q > 395$~yr.  However, it is not clear how long the source was in outburst when it was first detected.  If we assume that $t_0 = 17$~yr, then the lower limit for the quiescent time increases to $t_q > 560$~yr.  We note that the long outburst episode of this transient will have heated the crust significantly out of thermal equilibrium with the core and therefore the crust might still be cooling down toward equilibrium again. If that is the case, then the thermal flux related to the state of the core is even lower than the measured value (hence the core is cooler) and the quiescent episode should be even longer to allow the core to cool down.
We note that if CX2 is not the counterpart of the transient quiescent flux would be lower and therefore the predicted quiescent episodes would be even longer.

Disk-instability models suggest that such a long quiescent should not occur.  During quiescence the disk will slowly fill up again and at some point, even a small change in mass transfer rate should trigger an outburst \citep[e.g.,][]{lasota2001}.  Another neutron star X-ray transient observed to have been in outburst for $>$10 yr is KS 1731-260.  The quiescent flux of this neutron star also leads to the length of predicted quiescent episodes of several hundred years \citep{wijnands2001}.  However, the neutron star X-ray transient in the globular cluster NGC~6440 has much shorter outbursts and quiescent periods that are consistent with the observed quiescent flux \citep{cackett2005}. Thus, it appears that for the quasi-persistent neutron stars X1732$-$304 and KS 1731-260 to be as cool as observed in quiescence either the quiescent periods are several hundreds of years long, or the quiescent periods are much shorter and enhanced neutrino cooling is required for the neutron star to cool more quickly.  In the latter case it is possible that in these systems the neutron stars are relatively heavy as suggested by \citet{colpietal01}.

\subsection{Comparison with other globular clusters}

\citet{pooley2003} found a relationship between the number of X-ray sources (with 0.5-6 keV luminosity $L_X > 4\times10^{30}$ ergs s$^{-1}$) in a globular cluster and the stellar encounter rate of the cluster (see their Figure 2).  These authors calculate the stellar encounter rate by performing a volume integral of the encounter rate per unit volume, $R \propto \rho^2/v$ (where $\rho$ is the density and $v$ is the velocity dispersion), out to the half-mass radius.  This requires knowing the core density and velocity dispersion.  Unfortunately, for Terzan 1 there are no measurements of the velocity dispersion available.  However, we can relate the velocity dispersion to the core density, $\rho_0$, and core radius, $r_c$, via the virial theorem, and thus can estimate the encounter rate by $\rho_0^{1.5} r_c^2$ \citep{verbunt03}.  As Terzan 1 is a `core-collapse' cluster and the core is not well-defined any estimate for the encounter rate based on core values will be unreliable.  However, using the values for the core density, $\rho_0$, and core radius, $r_c$, from the Harris catalog \citep[][February 2003 version]{harris96} we find that the core is a factor of 3.4 less dense than that of 47 Tuc and the core radius is a factor of 8 times smaller.  We therefore estimate that the encounter rate of Terzan 1 will be a factor of $3.4^{1.5} \times 8^2 \sim 400$ smaller than in 47 Tuc.  Thus, based on the core density and core radius, the encounter rate for Terzan 1 is many tens to hundreds of times smaller than the value of 47 Tuc.  As 47 Tuc has a encounter rate of 396 (in the \citep{pooley2003} normalisation), we estimate that Terzan 1 has an encounter rate of $\sim 1$, which from the \citet{pooley2003} relationship would imply that there should only be $\sim1$ X-ray source in Terzan 1.  This is clearly not the case, as we detect 14 sources (though 2 are predicted to be not associated with the cluster).

As noted above, as Terzan 1 is a core-collapse cluster, therefore any estimate  based on the core radius and density will be unreliable. It is therefore important to determine whether any change in these parameters could make this cluster consistent with the relationship.  If the core radius is larger than assumed, this will have the corresponding effect of decreasing the computed density of the core, therefore it is difficult to increase the estimated encouter rate of Terzan 1 by the required factor of $\sim$50 to make it consistent with the \citet{pooley2003} relationship.  For instance, increasing the assumed core radius by a factor of 3 (to 7\farcs2) decreases the computed core density by a factor of 3.09 \citep[following ][]{Djorg93}, and thus will increase the computed encounter rate by only a factor of 1.66.  Uncertainty in the distance has a similarly small effect on the cluster parameters. Uncertainty in the reddening towards Terzan 1 can also have an effect on the estimated core density  as the core density is calculated from the extinction corrected central surface brightness.  In fact, using E(B-V)=2.48 \citep{ortolanietal99}, rather than the value from the Harris catalog, will increase $\rho_0$ by 77\%, although still not enough to eliminate the discrepancy with the \citet{pooley2003} relation.

To account for the number of detected X-ray sources it may be that the cluster was previously much larger and that most of the stars have been lost, perhaps due to passages through the Galactic disk as was suggested for NGC 6397 by \citet{pooley2003}. This idea is supported by the fact that Terzan 1 is in the bulge of the Galaxy and has the closest projection to the Galactic centre among known globular clusters.  Further evidence for a previously more massive cluster comes from the spectroscopic study of Terzan 1 by \citet{idiart02}.  These authors identify apparent members of Terzan 1 with substantially higher metallicity and suggest that Terzan 1 may have captured these stars from the bulge during a previous epoch when the cluster was significantly more massive. 

\section{Conclusions}

We have presented a $\sim$19 ks \textit{Chandra} ACIS-S observation of the globular cluster Terzan 1.  Within 1.4 arcmin of the cluster centre we detect 14 X-ray sources and predict that 2 of these are not associated with the cluster (background AGN or foreground objects).  The brightest of these sources, CX1, is consistent with the position of CXOGLB J173545.6-302900 first observed during a short \textit{Chandra} HRC-I observation of Terzan 1 \citep{WHG02}.  This source has a particulary hard spectrum for a globular cluster source, with a simple power-law fit giving a photon index of $0.2\pm0.2$.  Such a hard spectrum suggests that this could be an intermediate polar \citep{muno2004}.  The position of the second brightest source, CXOGLB J173547.2-302855 (CX2), is the only source to have a position that is consistent with all 3 of the previous {\it ROSAT} pointings that observed the neutron star transient X1732$-$304 \citep{JVH95} as well as the {\it EXOSAT} observation.  This source's X-ray color and spectrum suggest that it could be a quiescent neutron star and therefore we find that this source is the most likely candidate for the quiescent counterpart of X1732$-$304.  CX2 has a position closest to the position of the radio source detected in Terzan 1 by the VLA \citep{martietal98}, though the positions are not fully consistent.  Assuming standard core cooling, from the quiescent flux of CX2 we predict extremely long ($>$400 yr) quiescent episodes of X1732$-$304, or enhanced core cooling, to allow the neutron star to cool.  Having estimated the stellar encounter rate of this cluster we find significantly more sources than expected by the relationship of \citet{pooley2003} perhaps because the cluster was previously much larger and that most of the stars have been lost due to passages through the Galactic disk.

\subsection*{Acknowledgements}

The authors wish to thank the referee, Frank Verbunt, for careful and helpful comments which have improved this paper.  EMC gratefully acknowledges the support of a PPARC Studentship at the University of St Andrews.  DP gratefully acknowledges support provided by NASA through Chandra Postdoctoral Fellowship grant PF4-50035 awarded by the Chandra X-ray Center, which is operated by the Smithsonian Astrophysical Observatory for NASA under contract NAS8-03060.  The authors wish to thank P. Broos for his support with ACIS Extract.

This work makes use of the Digitized Sky Surveys which were produced
at the Space Telescope Science Institute under U.S. Government grant NAG W-2166.
This publication makes use of data products from the Two Micron All Sky Survey,
which is a joint project of the University of Massachusetts and the Infrared
Processing and Analysis Center/California Institute of Technology, funded by
the National Aeronautics and Space Administration and the National Science
Foundation. This research has made use of the USNOFS Image and Catalogue Archive operated by the United States Naval Observatory, Flagstaff Station (http://www.nofs.navy.mil/data/fchpix/).  This work has also made use of the VizieR online database of astronomical catalogues \citep{ochsenbein00}.

\bibliographystyle{mn2e}
\bibliography{iau_journals,qNS_mnras}

\end{document}